\documentstyle[preprint,prl,aps]{revtex}
\begin{document}
\draft
\title{Energy relaxation of an excited electron gas in 
quantum wires: many-body electron LO-phonon coupling}
\author{Lian Zheng and S. Das Sarma} 
\address{Department of Physics,
University of Maryland, College Park, Maryland 20742-4111}
\date{\today}
\maketitle
\begin{abstract}
We theoretically study energy relaxation via LO-phonon emission
in an excited one-dimensional electron gas
confined in a GaAs quantum wire structure.
We find that the inclusion of phonon renormalization effects
in the theory extends the LO-phonon dominated loss regime down 
to substantially
lower temperatures.
We  show that a simple  plasmon-pole approximation works well for 
this problem, and discuss implications of our results
for low temperature electron heating experiments
in quantum wires.
\end{abstract}
\narrowtext
When excess energy is supplied to an electron gas, 
either by a strong applied electric field or by
optical excitations, the electron gas becomes ``hot'',
{\it ie.}, it goes out of equilibrium with the lattice attaining
higher electron temperature than the ambient lattice temperature. 
A hot electron gas loses energy
to its surroundings in order to achieve equilibrium.  This energy 
loss process is usually accomplished through emission of phonons.  
In polar semiconductor materials such as GaAs, the electron 
LO-phonon Fr\"{o}hlich coupling is significantly
stronger than the electron acoustic-phonon coupling.
The most efficient energy relaxation process for the hot-electron gas,
except at very low temperatures,
is therefore to emit LO phonons.
The understanding  of this energy relaxation process
is of great technological
importance since actual electronic devices work mostly
under high-field hot-electron conditions. The study of this 
subject also constitutes a direct probe
of a fundamental interaction in condensed matter physics, namely,
the electron-phonon interaction. 
There has been considerable recent theoretical and experimental 
interest in the hot-electron energy relaxation problem 
in polar semiconductors,
particularly in three-dimensional (3D) and two-dimensional (2D) 
GaAs structures \cite{ed1,hd1,hd1p,ts1,hd2,hd2p,hd2pp,hd2ppp,tm1}. 
More recently, one-dimensional (1D) hot-electron relaxation in quantum wire 
structures has been considered theoretically
\cite{1d1,1d1p,1d1pp}, motivated 
by the fact that there has been successful growth of one-dimensional 
GaAs quantum-well wires with only the lowest subband occupied \cite{1d2}. 
In this article, 
we develop a many-body theory for hot-electron energy relaxation in one-dimensional quantum wires within the electron temperature model, 
taking full account of LO-phonon renormalization effects which have been 
left out of existing theories \cite{1d1,1d1p}.
Our results agree with existing theoretical results at high 
electron temperatures ($T>100K$),
but at lower temperatures we find significant contributions to the 
energy relaxation arising from renormalized LO-phonons which
have been left out of existing calculations.

It has been well recognized from the study of energy relaxation
in 2D and 3D systems \cite{hd1p} that it is important to include
the effect of the phonon propagator renormalization
at low temperatures.
This renormalization takes into consideration phonon self-energy 
due to electron-phonon 
interaction which opens up additional 
channels to energy relaxation 
from coupled electron-phonon excitations. 
These contributions
become increasingly important as the electron temperature 
$T$ decreases. 
At low enough temperatures $k_BT\ll\hbar w_{\rm LO}$, 
where $\omega_{\rm LO}$ is the frequency of the 
dispersionless LO phonon, the inclusion
of the phonon propagator renormalization may enhance the energy 
relaxation rate by orders of magnitude
compared with loss through bare LO-phonon (which is exponentially small
for $k_BT\ll\hbar\omega_{LO}$).
Deviation from the naive bare-phonon result,
which gives an exponentially decaying energy relaxation rate
as electron temperature is lowered, is a ubiquitous 
phenomenon in hot-electron energy loss experiments 
\cite {hd2,hd2p,hd2pp,hd2ppp,tm1},
and has usually been uncritically ascribed to acoustic-phonon contribution,
in spite of the fact that merely including the acoustic-phonon emission 
could not account for the total observed power loss. 
This puzzle 
is resolved by including the enhancement 
from the renormalization of the LO-phonon propagator. 
Theoretical calculations \cite{hd1p,ts1}
including the LO-phonon renormalization effects
show excellent qualitative and reasonable quantitative agreement 
with the existing experiments in 2D and 3D systems. 
The goal of the present work is to apply 
a similar theoretical model to study hot-electron energy relaxation
via coupling to bulk LO-phonons 
in semiconductor quantum wire structures. 
Our emphasis is on the low temperature regime,
where the many-body 
phonon propagator renormalization enhancement is important.
We compare the energy relaxation rate through the LO phonon coupling with 
that through acoustic phonon coupling, and estimate the 
electron temperature for the crossover from the LO-phonon  
dominated energy
relaxation to acoustic phonon dominated 
energy relaxation as the electron temperature decreases.  
As expected, we find that the enhancement of the energy relaxation 
from the LO-phonon propagator renormalization lowers the
crossover temperature substantially.  
For commonly available GaAs quantum wire 
carrier density of $n\sim10^5{\rm cm}^{-1}$,
the naive picture where the phonon renormalization is ignored 
would suggest a crossover temperature of $30K$, 
where acoustic phonon emission becomes comparable to LO-phonon emission,
while our estimated 
crossover temperature with the phonon renormalization included is
well below $30K$.
We also show that the plasmon-pole approximation gives reasonably 
accurate result for this problem, 
which is a direct consequence of the fact that 
in a 1D system
the plasmon-LO-phonon coupling is strong \cite{hs1}
for all carrier densities
and that quasiparticle electron-hole 
excitations are severely suppressed. 
The advantage of using the plasmon-pole approximation is
that the energy relaxation rate can be written as an analytic 
compact formula  which is formally as simple as 
the corresponding bare phonon result and yet it
gives quantitatively accurate result.

We adopt the standard electron-temperature model \cite{ed1,tm1}
which assumes
that the electron gas and the substrate lattice are separately 
in equilibrium at different temperature $T$ and $T_L$
with $T>T_L$.
We also assume that the electrons in the quantum wire 
structure occupy only the lowest subband, which is 
a valid assumption for carrier density $n\leq10^6{\rm cm}^{-1}$
and wire width $a\leq300\AA$.
For lattice temperature $T_L=0K$, the energy relaxation rate is \cite{hd1p}
\begin{equation}
P=\sum_q\int{d\omega\over\pi}\omega n_T(\omega)
|M_q|^2{\rm Im}\chi^{\rm ret}(q,\omega){\rm Im}D^{\rm ret}(q,\omega),
\label{equ:p1}
\end{equation}
where $n_T(\omega)$ is Bose distribution factor at temperature $T$.
$\chi^{\rm ret}(q,\omega)$ is the retarded density-density 
response function
for an uncoupled interacting 1D electron gas. 
In the random-phase approximation (RPA),
$\chi(q,\omega)=\chi_o(q,\omega)/[1-v_q\chi_o(q,\omega)]$, 
with the finite temperature $\chi_o(q,\omega)$ 
of a 1D free electron gas obtained 
from its zero-temperature counterpart by an 
integration over chemical potential \cite{hus1}.
For a quantum wire with finite wire widths, 
the Coulomb interaction potential is 
$v_q=(e^2/\epsilon_\infty)\int d\eta |I(\eta)|^2H(q,\eta)|
/(q^2+\eta^2)^{1/2}$. The form factor $I(\eta)$ and 
$H(q,\eta)$ used in our calculation are taken from  
the infinite well confinement model \cite{1d1p}.
The Fr\"{o}hlich coupling matrix is 
$|M_q|^2=v_q(\omega_{\rm LO}/2)[1-(\epsilon_\infty/\epsilon_0)]$,
with $\epsilon_0 (\epsilon_\infty)$ as the static (high frequency) 
dielectric constant.
The phonon propagator in Eq. (\ref{equ:p1}) is 
\begin{equation}
D(q,\omega)={2\omega_{\rm LO}\over\omega^2-\omega^2_{\rm LO}
-2\omega_{\rm LO}|M_q|^2\chi(q,\omega)}. 
\label{equ:d1}
\end{equation}

The last term in the denominator is the phonon self-energy.
When this phonon self-energy is ignored, one has 
Im$D^{\rm ret}(q,\omega)=$Im$D^{\rm ret}_0(q,\omega)
=\pi[\delta(\omega+\omega_{\rm LO})-\delta(\omega-\omega_{\rm LO})]$.
Inserting this into Eq. (\ref{equ:p1}), one obtains the energy 
relaxation rate for bare phonon emission as
\begin{equation}
P_0=\omega_{\rm LO}n_T(\omega_{\rm LO})\sum_q
(-2)|M_q|^2\chi(q,\omega_{\rm LO}).
\label{equ:po1}
\end{equation}
The characteristic of the bare phonon result is an approximate  
exponential temperature dependence $P_0\propto{\rm exp}(-\omega
_{\rm LO}/k_BT)$, which comes from the Bose factor $n_T(\omega_{\rm LO})$. 

With a standard plasmon-pole approximation \cite{hd1p} 
for $\chi(q,\omega)$, the spectral function of the phonon 
propagator in Eq. (\ref{equ:d1}) becomes a pair of 
$\delta-$functions at the coupled plasmon-phonon 
excitation frequencies.
Inserting this spectral function into Eq. (\ref{equ:p1}), a compact
expression for energy relaxation rate is obtained as
\begin{eqnarray}
&&P_{\rm PP}=P_++P_-
\nonumber \\
&&P_{\pm}=\sum_q\omega_\pm n_T(\omega_\pm)
{\omega_{\rm LO}|\omega_\pm^2-\omega_P^2|\over
\omega_\pm(\omega_+^2-\omega_-^2)}|M_q|^2
(-2){\rm Im}\chi(q,\omega_\pm),
\label{equ:tt}
\end{eqnarray}
where $\omega_\pm$ are the coupled plasmon-phonon 
excitation energies \cite{hd1p}, 
and $\omega_P$ is the uncoupled plasmon excitation for a 
1D electron gas \cite{hus1,ls1}.
Note that we use the plasmon-pole approximated form of $\chi(q,w)$ 
only in the phonon self-energy in Eq. (\ref{equ:d1}), not
in the energy loss rate of Eq. (\ref{equ:p1}).
Most of our calculations are concerned with comparing 
the energy relaxation rates from the different approximations
expressed in Eq. (\ref{equ:p1}), (\ref{equ:po1}), 
and (\ref{equ:tt}).

In our calculation, the hot phonon ``bottleneck'' 
\cite{ed1,hd1p,bn1} effect is ignored.  
This is equivalent to setting the lifetimes of the emitted LO-phonons 
to zero. 
The reasons for doing this are that the effect from the finite
lifetime of the emitted phonon is less important
at low temperatures and the values of the lifetimes for
the electron-phonon coupled excitations are unknown.
It is easy to include the hot phonon bottleneck effect in the theory 
\cite {hd1p} if the phonon lifetime is known.

The results of our calculation
are shown in Fig. \ref{f1} to \ref{f4}, where we take the parameters which 
are appropriate for GaAs materials:
$\epsilon_0=12.9,\ \epsilon_\infty=10.9$, and $\omega_{\rm LO}=36.8$
meV. In Fig. \ref{f1}, energy relaxation rates with
bare phonons, renormalized phonons, and plasmon-pole approximation
are shown as functions of electron temperature. 
The bare phonon result shows an approximate exponential 
temperature dependence mainly due to the Bose factor $n_T(\omega_{\rm LO})$.
In addition to phonon-like excitation, 
the renormalized phonon propagator also contains coupled
plasmon- and quasiparticle-
like excitations. The phonon-like excitation
has large spectral weight and  high energy ($\hbar\omega_{\rm LO}=427K$),
while the 
plasmon- and quasiparticle-like excitations have small spectral weights,
but with arbitrarily low energies. 
At high temperatures ($k_BT\sim\hbar\omega_{\rm LO}$),
the phonon-like excitation dominates
because of its large spectral weight. The energy relaxation rate
of the renormalized phonon is essentially the same as that 
of bare phonon. As the temperature is lowered, the plasmon-
and quasiparticle-like excitations begin to dominate
because of their low energies. The energy
relaxation rate of the renormalized phonon starts to deviate 
from that of the bare phonon.
As shown in Fig. \ref{f1}, the low temperature enhancement
to the energy relaxation from the phonon renormalization
at high density ($n=10^6{\rm cm}^{-1}$) is weaker than that 
at low density ($n=10^5{\rm cm}^{-1}$),
but it is still significant.
The temperature where the deviation from the bare phonon 
result begins is about $100K$ (compare to $\hbar\omega_{\rm LO}=427K$).
Due to strong plasmon LO-phonon coupling 
\cite{hs1} and phase space restriction
on quasiparticle excitation in 1D systems,
plasmon-pole approximation is expected to give reasonably accurate
result. This is basically true in Fig. \ref{f1}, especially
for the case of low density ($n=10^5{\rm cm}^{-1}$). 

In a polar semiconductor material such as GaAs, electron LO-phonon 
coupling is much stronger than electron-acoustic phonon coupling.
Hot-electron energy relaxation through the LO-phonon channel 
dominates over that through the acoustic phonon channel,  
except at low temperatures where emission of the high energy 
LO-phonon is effectively frozen by energy conservation. 
The temperature for the crossover 
from the LO-phonon dominated energy relaxation
to the acoustic phonon dominated energy relaxation should be 
significantly shifted by the enhancement to the energy relaxation
from LO-phonon renormalization. 
In Fig. \ref{f2}, the energy relaxation rates for
bare LO-phonons, renormalized LO-phonons and 
deformation-potential acoustic phonons \cite{1d1pp} are
shown as functions of electron temperature $T$.
The crossover temperature without considering the LO-phonon 
renormalization is about $30K$ for both densities. 
When the LO-phonon renormalization effect is included, 
the crossover temperature is shifted down to about $10K$ for
electron density of $n=10^6{\rm cm}^{-1}$ and 
well below $1K$ for $n=10^5{\rm cm}^{-1}$. 
The basic conclusion from our calculation 
is that the crossover temperature
can be shifted down substantially by the phonon renormalization effect.

In Fig. \ref{f3}, energy relaxation rates as functions 
of electron density and wire widths are shown respectively. 
As the density increases, among other things, the
screening is increased and the energies of the coupled electron-phonon
excitation are raised,  so the energy relaxation rate for 
bare phonons declines slightly while the energy relaxation rate 
for renormalized phonons decreases more noticeably.
The dependence of energy relaxation
rate on wire widths is shown to be weak.

Finally, the energy resolved contribution 
to the relaxation rate as functions of renormalized phonon frequency
is shown in Fig. \ref{f4}, 
where $I(\omega)$ is defined by rewriting Eq. (\ref{equ:p1})
as $P=\int_0^\infty I(\omega)d\omega$.
It is clearly seen that at high temperature ($T=250K$) the energy 
relaxation is dominated by emission of bare LO-phonons with 
$\omega\sim\omega_{LO}$,
while at low temperature ($T=36K$), energy relaxation comes mainly from
emission of renormalized phonons
with $\omega\ll\omega_{LO}$.  
The significant contribution to energy relaxation 
from low energy phonon emission at $T=36K$ suggests that 
it is necessary to incorporate many-body phonon renormalization
into energy relaxation theories in order to obtain
meaningful results at low temperatures.

This work is supported by the U.S.-ONR and the
U.S.-ARO.

\begin{figure}
\caption{ 
Energy relaxation rate per electron as functions of 
electron temperature $T$. The solid-line, dot-line, 
and dot-dashed-line are respectively the results with 
renormalized phonon propagator, bare phonon propagator,
and plasmon-pole approximated phonon propagator. 
The widths of the quantum wire are $L_y=L_z=200\AA$.
(a) electron density $n=10^5{\rm cm}^{-1}$. 
(b) electron density $n=10^6{\rm cm}^{-1}$.  
}
\label{f1}
\end{figure}

\begin{figure}
\caption{ 
Energy relaxation rate per electron as functions of
electron temperature $T$. The solid-line and dot-line
are respectively the results with
renormalized LO phonon propagator and bare LO phonon propagator.
The dot-dashed-line is the energy relaxation rate through
acoustic phonon coupling (see Ref. 12).
The widths of the quantum wire are $L_y=L_z=200\AA$.
(a) electron density $n=10^5{\rm cm}^{-1}$.
(b) electron density $n=10^6{\rm cm}^{-1}$.
}
\label{f2}
\end{figure}

\begin{figure}
\caption{ 
(a) Energy relaxation rate per electron as functions of
electron density $n$. 
The widths of the quantum wire are $L_y=L_z=50\AA$.
(b) Energy relaxation rate per electron as functions of
quantum wire widths. 
The electron density is $n=10^5{\rm cm}^{-1}$.
In both (a) and (b), the solid-line and dot-line
are respectively the results of
renormalized phonon and bare phonon,
and the electron temperature is $T=35K$.
}
\label{f3}
\end{figure}

\begin{figure}
\caption{Energy relaxation rate as functions of renormalized phonon
frequency $\omega$ at high temperature (circle-marked line) and low 
temperature (diamond-marked line), where $I(\omega)$ is defined 
in the text. The diamond-marked line represents the low temperature 
result multiplied by a factor of $10^4$. 
The widths of the quantum wire are $L_y=L_z=50\AA$.
The electron density is $n=10^6{\rm cm}^{-1}$.
}
\label{f4}
\end{figure}

\end{document}